\documentstyle[psfig]{mn}
\begin{document}

\def\aj{AJ}                   
\def\araa{ARA\&A}             
\def\apj{ApJ}                 
\def\apjl{ApJ}                
\def\apjs{ApJS}               
\def\ao{Appl.~Opt.}           
\def\apss{Ap\&SS}             
\def\aap{ A\&A}                
\def\aapr{ A\&A~Rev.}          
\def\aaps{ A\&AS}              
\def\azh{ AZh}                 
\def\baas{ BAAS}               
\def\jrasc{ JRASC}             
\def\memras{ MmRAS}            
\def\mnras{ MNRAS}             
\def\pra{ Phys.~Rev.~A}        
\def\prb{ Phys.~Rev.~B}        
\def\prc{ Phys.~Rev.~C}        
\def\prd{ Phys.~Rev.~D}        
\def\pre{ Phys.~Rev.~E}        
\def\prl{ Phys.~Rev.~Lett.}    
\def\pasp{ PASP}               
\def\pasj{ PASJ}               
\def\qjras{ QJRAS}             
\def\skytel{ S\&T}             
\def\solphys{ Sol.~Phys.}      
\def\sovast{ Soviet~Ast.}      
\def\ssr{ Space~Sci.~Rev.}     
\def\zap{ ZAp}                 
\def\nat{ Nature }              
\def\iaucirc{ IAU~Circ.}
\def\aplett{ Astrophys.~Lett.}
\def\apspr{ Astrophys.~Space~Phys.~Res.}
\def\bain{ Bull.~Astron.~Inst.~Netherlands}
\def\fcp{ Fund.~Cosmic~Phys.}
\def\gca{ Geochim.~Cosmochim.~Acta}
\def\grl{ Geophys.~Res.~Lett.}
\def\jcp{ J.~Chem.~Phys.}      
\def\jgr{ J.~Geophys.~Res.}    
\def\jqsrt{ J.~Quant.~Spec.~Radiat.~Transf.}
\def\memsai{ Mem.~Soc.~Astron.~Italiana}
\def\nphysa{ Nucl.~Phys.~A}
\def\physrep{ Phys.~Rep.}
\def\physscr{ Phys.~Scr}
\def\planss{ Planet.~Space~Sci.}
\def\procspie{ Proc.~SPIE}
\let\astap=\aap
\let\apjlett=\apjl
\let\apjsupp=\apjs
\let\applopt=\ao
\def\phn{\phantom{0}}
\def\phd{\phantom{.}}
\def\phs{\phantom{$-$}}
\def\phm#1{\phantom{#1}}
\def\sun{\hbox{$\odot$}}
\def\earth{\hbox{$\oplus$}}
\def\lesssim{\mathrel{\hbox{\rlap{\hbox{\lower4pt\hbox{$\sim$}}}\hbox{$<$}}}}
\def\gtrsim{\mathrel{\hbox{\rlap{\hbox{\lower4pt\hbox{$\sim$}}}\hbox{$>$}}}}
\def\sq{\hbox{\rlap{$\sqcap$}$\sqcup$}}
\def\arcdeg{\hbox{$^\circ$}}
\def\arcmin{\hbox{$^\prime$}}
\def\farcs{\hbox{$.\!\!^{\prime\prime}$}}       
\def\arcs{\hbox{$^{\prime\prime}$}}       
\def\deg{$^\circ$}
\def\kms{$\rm km\;s^{-1}$}
\def\Msun{{\rm M}_\odot}
\def\Lsun{{\rm L}_\odot}

\title{Evidence for a Massive Dark Object in NGC~4350} 
\author[E. Pignatelli, P. Salucci and L. Danese]{Ezio Pignatelli\thanks{e-mail: pignatel@sissa.it}, 
Paolo Salucci, and Luigi Danese \\ 
SISSA, via Beirut 4, I-34014 Trieste, Italy}

\maketitle
\begin{abstract}

In this work we build a detailed dynamic model for a S0
galaxy possibly hosting a central massive dark object
(MDO). We show that the photometric profiles and the
kinematics along the major and minor axes, including the
$h_3$ and $h_4$ profiles, imply the presence of a central
MDO of mass \hbox{$M_{\rm MDO} \sim 1.5 - 9.7 \cdot
10^{8} \Msun$}, i.e. $0.3-2.8$\% of the mass derived for
the stellar spheroidal component.  Models without MDO are
unable to reproduce the kinematic properties of the inner
stars and of the rapidly rotating nuclear gas.

The stellar population comprise of an exponential disc
(27\% of the light) and a diffuse spheroidal component
(73\% of the light) that cannot be represented by a
simple de Vaucouleurs profile at any radius. The $M/L$
ratios we found for the stellar components (respectively
3.3 and 6.6) are typical of those of disc and elliptical
galaxies.

\end{abstract} 

\section{Introduction} 
\label{sec:intro}

There is increasing evidence that most galaxies with a
large spheroidal component host a Massive Dark Object
(MDO) at their centre, with masses ranging from $\sim
10^8 \Msun$ to $2 \cdot 10^{10} \Msun$ \cite{ho99}.

This evidence was first reviewed by Kormendy \& Richstone
\shortcite{KR95}, suggesting that in kinematically hot
galaxies (ellipticals and S0s) the mass of the central
object $M_{\rm MDO}$ correlates with the mass of the hot
stellar component $M_{\rm sph}$. For the ratio $x \equiv
\log (M_{\rm MDO}/M_{\rm sph})$ they found a Gaussian
distribution with average value $x=-2.5$ and variance
$\sigma=0.2$.  More recently, assuming isotropy in the
velocity dispersion tensor, Magorrian et
al. \shortcite{M98} exploited the high resolution of HST
photometry and ground based spectroscopy to estimate the
MDO mass of 36 E/S0s, finding a similar correlation $x =
-2.28 \pm 0.5$.

Exploiting the data from a sample of 30 galaxies, for
which at least two independent MDO mass estimates are
available, Salucci et al. \shortcite{BH1} found an
average value $x=-2.60 \pm 0.3$, quite close to that
obtained by Ho et al.  \shortcite{ho99} and Ford et
al. \shortcite{ford99}. This result has been used to
determine the mass function of MDOs hosted in spheroidal
galaxies and turns out to be fully consistent with the
mass function of accreted matter required to fuel QSO
activity.  This supports the idea that MDOs are the
remnant BHs of a past era of nuclear galactic activity.

On the other hand, using several hundred rotation curves,
Salucci et al. \shortcite{BH2} recently put stringent
upper limits on the mass of the MDOs resident in
late-type galaxies, showing that they must be on average
at least 10-100 times smaller than the ones claimed for
ellipticals. If one assumes that MDOs are the remnants of
past QSO activity, this implies that the contribution to
the shining phase of quasars by BHs hosted in late-type
spirals is totally negligible.

Finally, recent results \cite{mbhsigma1,mbhsigma2} showed
that the central MDO mass strongly correlates with the
galaxy velocity dispersion. In particular Gebhardt et
al. \shortcite{mbhsigma2} pointed out that ellipticals
and bulges fall again in a fundamental plane when they
are located in a four-dimension plane with coordinates
($M_{BH}-\log L-\log\sigma,\log R_e$). This property
implies some fundamental connection between the black
hole and the bulge.

In this context it is interesting to investigate the role
of S0 and early-type disc galaxies. In view of their
massive bulge, they must be considered as a possible
location for QSO remnants. However, the search for
central massive objects in S0 is hampered, as: (1) the
bulge/disc photometric decomposition is often uncertain;
(2) these objects have weak $H\alpha$ lines and,
consequently, the rotation curve inside the central 0.5
kpc cannot be accurately determined; (3) the rotation
curve is too disturbed by the strong asymmetric drift
effects to be representative of the circular velocity.

Detection of MDOs in this kind of galaxy requires
kinematic data of the innermost regions, and a detailed
dynamic model. This approach has proved effective in
detecting massive central dark objects in a few S0/Sa
galaxies, including NGC~4594 \cite{kormendy96}, M31
\cite{tsvetanov}, NGC~3115 \cite{emsellem99}, and
NGC~4342 \cite{cretton99}. Given the highly compact
stellar structure, central MDOs can only be detected
through accurate reconstruction of the galactic
components, including the bulge/disc decomposition.

The goal of this work is to investigate the possible
presence of an MDO in the nucleus of the S0 galaxy
NGC~4350, by applying the dynamical model of Pignatelli
\& Galletta \shortcite{me} to recent data. This work has
been set out as follows: in Section~\ref{sec:data} we
present the available data; in Section~\ref{sec:model} we
give a brief summary of the hypothesis and
characteristics of the model; in
Section~\ref{sec:results} we derive the best-fit
parameters of the model; Section.~\ref{sec:discuss} is
devoted to the discussion of the results and conclusions.

In this paper we assume a Hubble constant $H_0=75$ km
s$^{-1}$ Mpc$^{-1}$.

\section{The data}
\label{sec:data}

NGC~4350 is classified as a classical S0 galaxy in the
{\em Nearby galaxy catalog} (Tully 1988; see
Table~\ref{tab:start}).  It is observed almost edge-on
(from its axial ratio one can infer an inclination angle
$i>62^\circ$).

For the photometry, we adopt the R-band surface
brightness profiles published by Seifert and Scorza
\shortcite{scorza}.  According to these authors, NGC 4350
can be decomposed into disc and bulge components. The
disc is probably exponential, significantly decreasing
beyond 40 arcsec; the bulge on the other hand dominates
even in the outer parts of the galaxy, as indicated by
the substantial drop in the ellipticity profile. No
ring-like component, or inner stellar disc has been
detected.

\begin{table}
\begin{center}
\caption{Properties of NGC~4350 \protect\cite{ngc}. 
(1) Hubble type $T$;
(2) absolute blue magnitude $B_T$; 
(3) total blue luminosity $L_B$; 
(4) position angle P.A.; 
(5) observed apparent axial ratio at $D_{25}$; 
(6) distance $D$; 
(7) conversion scale from arcsec to parsec; 
(8) observed diameter at 25 mag/arcsec$^2$ isophote in blue.}
\label{tab:start}
\begin{tabular}{cccccccc}
\hline
\multicolumn{1}{c}{(1)} &
\multicolumn{1}{c}{(2)} &
\multicolumn{1}{c}{(3)} &
\multicolumn{1}{c}{(4)} &
\multicolumn{1}{c}{(5)} &
\multicolumn{1}{c}{(6)} &
\multicolumn{1}{c}{(7)} &
\multicolumn{1}{c}{(8)} \\
\multicolumn{1}{c}{Type} &
\multicolumn{1}{c}{$B_{\rm T}$} &
\multicolumn{1}{c}{$L_{\rm B}$} &
\multicolumn{1}{c}{P.A.} &
\multicolumn{1}{c}{(b/a)} &
\multicolumn{1}{c}{$D$} &
\multicolumn{1}{c}{scale} &
\multicolumn{1}{c}{$D_{25}$} \\
\multicolumn{1}{c}{[T]}&
\multicolumn{1}{c}{[mag]} &
\multicolumn{1}{c}{[$\Msun$]} &
\multicolumn{1}{c}{[\deg]} &
&
\multicolumn{1}{c}{[Mpc]} &
\multicolumn{1}{c}{[pc/$\prime\prime$]} &
\multicolumn{1}{c}{[$\arcmin$]} \\ 
\hline
-2 &11.94&$8\cdot10^9$&28&0.62&16.8&81.5&3.0\\
\hline
\end{tabular} 
\end{center}
\end{table}

For the kinematics, we adopt Fisher's detailed stellar
kinematical observations along the major and minor axis
\cite{fisher97}. These include the rotational velocity,
the velocity dispersion profile, and the profiles of the
Gauss-Hermite higher-order moments $h_3$, $h_4$
\cite{vdm93}. Moreover, they also include emission-line
gas kinematics of a rapidly rotating nuclear disc.  The
total number of indipendent measurements to be
simultaneously fitted by a model is 280.

The high rotational velocity of the nuclear gaseous disc, the
rapid increase of the stellar rotation curve in the first
$2''$ ($\approx 160$ pc) and the corresponding high peak
in the velocity dispersion profile suggest the presence
of a central concentration of mass.
 
Loyer et al. \shortcite{simien98} modeled this galaxy in
some detail using a Multi-Gaussian Expansion method. In
this paper, the photometric and kinematic data are
reproduced with a different method and the subsequent
analysis is extended to include the gas rotational
velocity and the profiles of the $h_3$, $h_4$
parameters. As pointed out by several authors
\cite{vdm98,cretton99,ngc3379}, the use of the higher
moments of the line-of-sight velocity distribution is
very relevant in determining the orbital anisotropies,
thus reducing the uncertainties on the MDO mass.

\section{The Galaxy Model}
\label{sec:model}

In order to investigate the gas and stellar kinematics we
use the self-consistent dynamical technique of Pignatelli
and Galletta \shortcite{me}.  We give here a brief
summary of the technique and the general assumptions
made.

A galaxy can be described by superposition of different
components.  For each component, we separately assume:
\begin{description}
\item[-] the density distribution is oblate, without triaxial structures;
\item[-] the isodensity surfaces are similar concentric spheroids;
\item[-] the surface density profile follows a simple $r^{1/4}$ or an exponential law;
\item[-] the local velocity distribution is Gaussian;
\item[-] the velocity dispersion is isotropic 
         ($\sigma_r = \sigma_\theta =\sigma_z$);
\item[-] the mass-luminosity ratio is constant with radius;
\end{description}
Our model does not consider the possible presence of triaxial
structures (bar; triaxial bulge; tilted component; warp) or of
anisotropy in the velocity distribution (with predominance of radial
or tangential orbits).

The model has $4n+2$ free parameters, where $n$ is the
number of adopted components: namely the luminosity,
scale length, mass-luminosity ratio and flattening
$\{L_{tot},\ r_e,\ M/L,\ b/a \}$ of each component plus
the inclination angle of the galaxy and the mass of the
central massive dark object $M_{MDO}$. In principle,
photometry can be used to constrain all these parameters
except the M/L ratios and the value of $M_{MDO}$, which
must be derived through the kinematics.

\section{Comparison between model and data} 
\label{sec:results}

\begin{figure*}
\psfig{figure=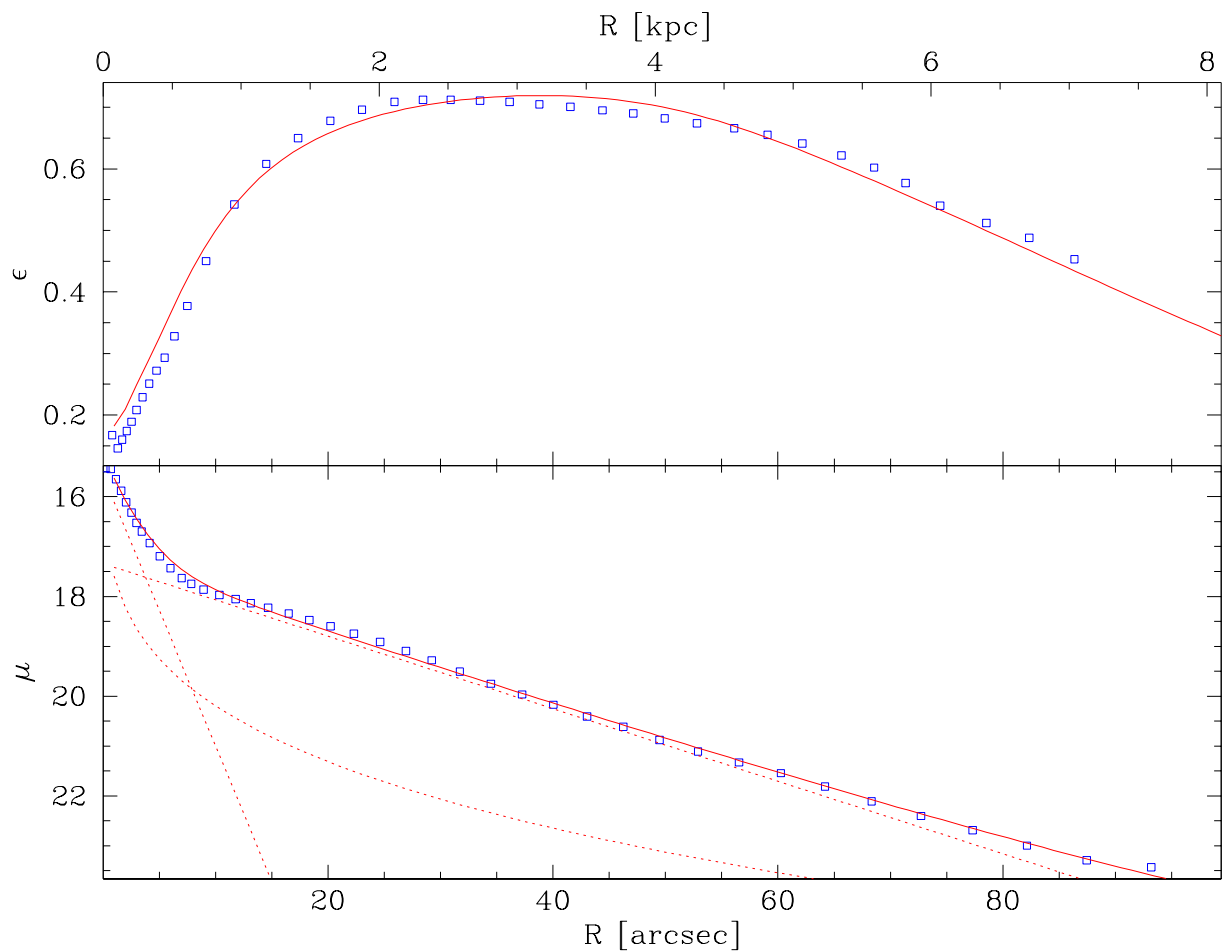,width=\textwidth,angle=270}
\caption{Photometric decomposition of NGC 4350. The
parameters of the different components are shown in Table
\protect\ref{tab:photo}. Data are taken from Seifert and
Scorza \protect\shortcite{scorza}. The dotted lines
represent the separate contributions of the three
different components to the global luminosity profile.}
\label{fig:photo}
\end{figure*}

\begin{table*}
\begin{center}
\begin{tabular}{lrrrcccccrccccc}
\hline
\noalign{\smallskip}
\multicolumn{3}{c}{scale radius}        &
\multicolumn{1}{c}{}                    &
\multicolumn{3}{c}{axial ratio}         &
\multicolumn{1}{c}{}                    &
\multicolumn{3}{c}{Luminosity}          &
\multicolumn{1}{c}{}                    &
\multicolumn{1}{c}{} \\ 
   \cline{1-3} \cline{5-7}  \cline{9-11}
\vspace*{-.35cm}                 \\
\multicolumn{1}{c}{r$_b$}               &
\multicolumn{1}{c}{r$_d$}               &
\multicolumn{1}{c}{r$_{h}$}      &
\multicolumn{1}{c}{}                    &
\multicolumn{1}{c}{(b/a)$_b$}           &
\multicolumn{1}{c}{(b/a)$_d$}           &
\multicolumn{1}{c}{(b/a)$_{h}$}  &
\multicolumn{1}{c}{}                    &
\multicolumn{1}{c}{L$_b$}               &
\multicolumn{1}{c}{L$_d$}               &
\multicolumn{1}{c}{L$_{h}$}      &
\multicolumn{1}{c}{}                    &
\multicolumn{1}{c}{i} 
                                        \\
\noalign{\smallskip}
\noalign{\smallskip}
$2 \pm 0\farcs1$ &$15\pm 0\farcs2 $ &$ 40\pm 6''$ &&  
$1. \pm 0.03$  &$  0.1 \pm 0.01$  &$ 0.95 \pm 0.06$  &&
$24.5 \pm 4\%$ &$26.5 \pm 1.5 \% $& $49 \pm 5 \%$&&85$\pm$5\deg\\
\hline
\end{tabular} 
\caption{Parameters of the decomposition shown in 
         Fig.~\protect\ref{fig:photo}. The errors define
         the 1$\sigma$ level in the parameter space on the
         basis of a reduced $\chi$-square analysis.}
\label{tab:photo}
\end{center}
\end{table*}

\subsection{Photometry}
 
Seifert and Scorza \shortcite{scorza} have performed
detailed CCD photometry in the R-band for a sample of 16
early-type disc galaxies, including NGC~4350. For each
galaxy, they determined the radial profiles for surface
brightness and ellipticity, together with a bulge/disc
decomposition without special assumptions on the radial
density profiles of the two components.  According to the
authors, the disc of NGC~4350 extends up to the innermost
regions of the galaxy and follows approximately an
exponential law; the spherical component dominates both
in the nucleus and in the outer parts, as indicated by
the strong external decrease in the ellipticity profile
(see Fig.~\ref{fig:photo}).

In order to describe the photometry of this galaxy we
have adopted a three-component model:
\begin{enumerate}
\renewcommand{\theenumi}{(\arabic{enumi})}
\item an almost edge-on exponential disc component; 
\item a diffuse spheroidal component, with a deVaucouleur surface 
  brightness profile, extending well beyond the disc
  itself, identified hereafter as ``stellar halo'';
\item a second spheroidal component, very compact and 
  with an {\em exponential} profile in the inner 2\arcs\ 
  (see Fig.~\ref{fig:photo}), which we identify as ``bulge''.
\end{enumerate}

The parameters derived by performing a best fit to the
photometric data are presented in Table~\ref{tab:photo},
together with their statistical uncertainties.

The rapid increase in the ellipticity from very small
values to a maximum of $\varepsilon \simeq 0.7$ at
$r\approx 30''$ (see Figure \ref{fig:photo}) and its slow
decline in the outer parts can be explained if at least
three components are used.  The two ``spheroidal''
components may represent two physically different stellar
systems, or, since the best-fit flattenings are very
similar, a single, almost spherical component with a
density profile not represented by the de Vaucoleur's law
(see discussion in the Section~\ref{sec:stellar}).

\subsection{Kinematics}
\label{par:kine}

\begin{figure*}
\centerline{\psfig{figure=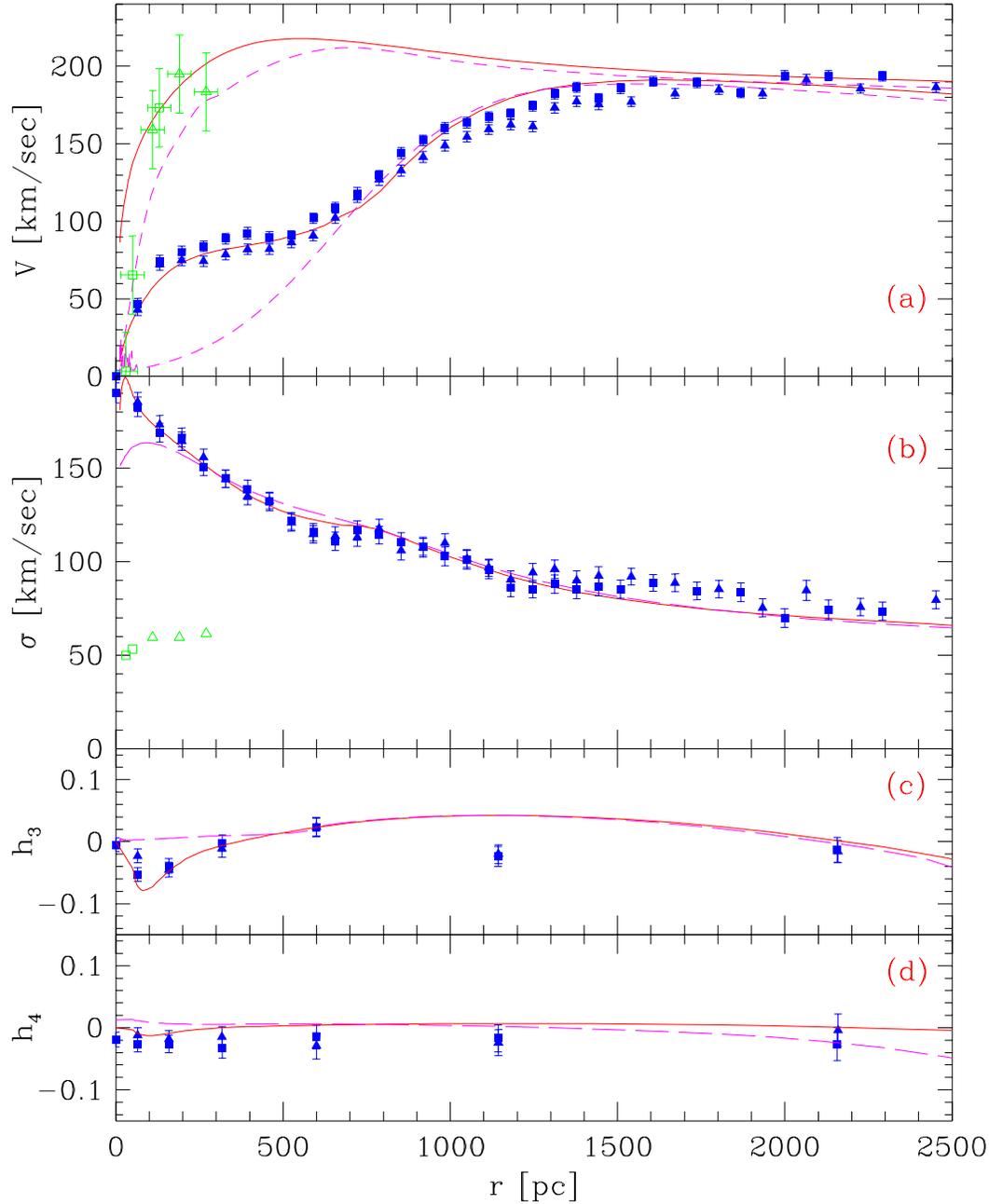,width=0.8\textwidth}}
\caption{Data from Fisher (1997) for the major axis,
compared with our best-fit models with ({\em solid line})
and without ({\em long-dashed line}) a central MDO.  The
squares and triangles represent data respectively from
the approaching and the receding side.  From the top: (a)
stellar ({\em filled symbols}) and gas ({\em open
symbols}) rotational velocities. The latter is also
compared with the circular velocity inferred from the two
models; (b) velocity dispersion profile; (c) $h_3$ radial
profile; (d) $h_4$ radial profile.  Data beyond 2.5 kpc
are not shown because they were not included in the fit.}
\label{fig:major}
\end{figure*}

\begin{figure*}
\centerline{\psfig{figure=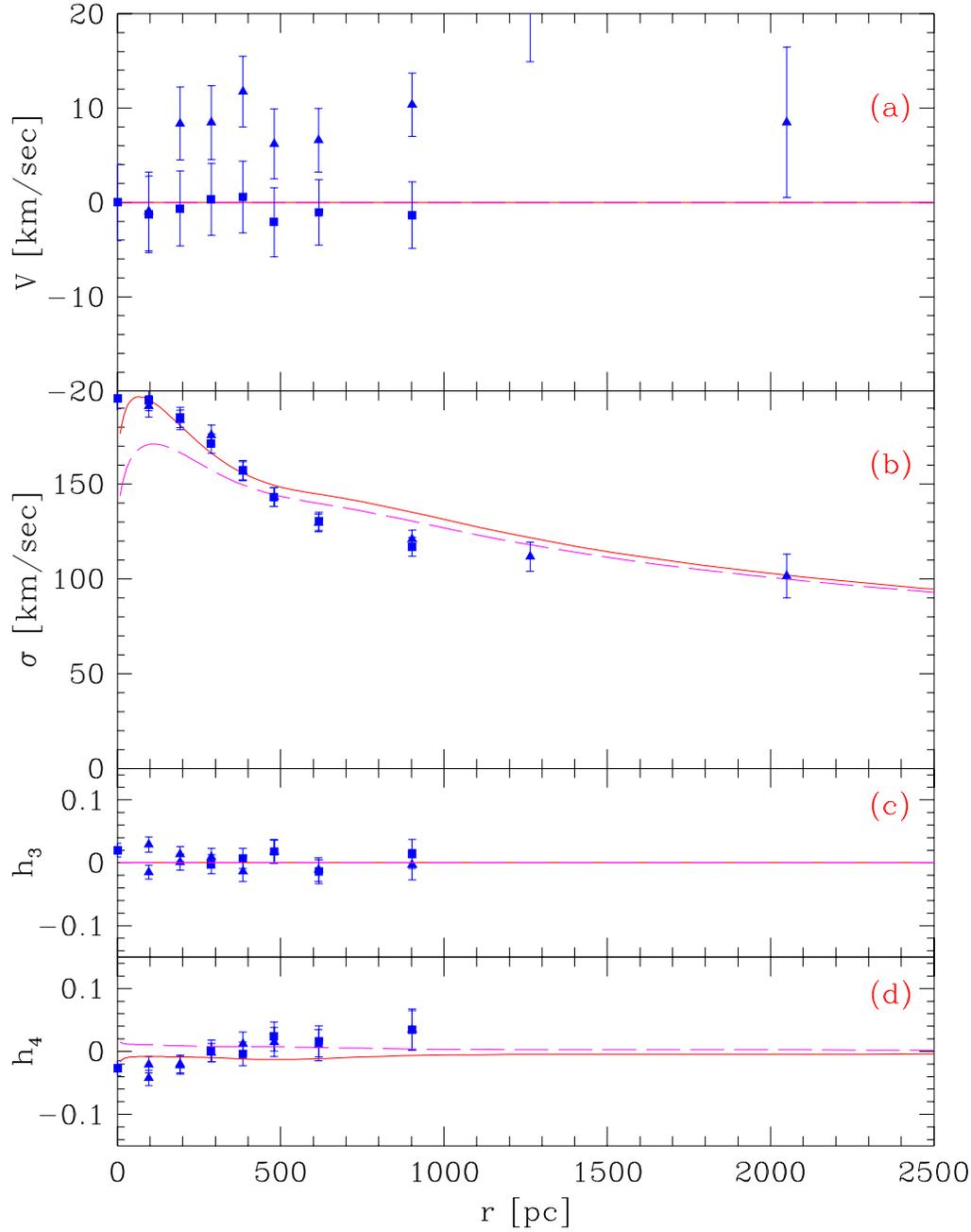,width=0.8\textwidth}}
\caption{Data from Fisher (1997) for the minor axis,
compared with our best-fit model with ({\em solid line})
and without ({\em long-dashed line}) a central massive
dark object. The notation used for the symbols is the
same as in Fig.~\protect\ref{fig:major}. }
\label{fig:minor}
\end{figure*}

Fisher \shortcite{fisher97} has published ground-based
kinematics of 18 early-type galaxies, including
NGC~4350. The data includes the stellar rotation curve,
velocity dispersion profile and the line-of-sight
velocity distribution described in terms of the
Gauss-Hermite expansion series terms $h_3$ and $h_4$
\cite{vdm93}.  Emission line spectroscopy reveals a disc
of rapidly rotating gas within the inner 3\arcs, which
appears kinematically decoupled from the stellar
component. In fact, while the gas is likely to follow
circular orbits, the stellar velocity is obtained by
taking into account the asymmetric drift effect as
\begin{equation}
V_{\rm rot}^2=V_{\rm circ}^2-\Delta V^2_{\rm asymdrift}, 
\label{eq:asymdrift}
\end{equation}
where the $ \Delta V^2_{\rm asymdrift}$ \cite{bible} is
essentially a function of the velocity dispersion and is
only slightly dependent on the stellar distribution. The
data are shown in Figs.~\ref{fig:major}-\ref{fig:minor}.

Most parameters of the model have been fixed by fitting
the photometry; the kinematic data are fitted by varying
only the $M/L$ ratios of the different stellar components
and the value of the MDO mass.

A robust upper limit to the central mass of the MDO can
be obtained by assuming that the gas follows a Keplerian
rotation curve. After convolution with seeing and slit
effects, we obtain $ M_{MDO}\lesssim 1.5\cdot 10^9
\Msun$. However, since part of the gravitational force is
due to the diffuse stellar component, we allowed the
value of the dark mass concentration to vary between zero
and $M_{max}=1.5\cdot 10^9 \Msun$.  The value of the
$M/L$ ratios for the spheroidal components was allowed to
vary from 1 to 10 times the corresponding value for the
disc.

We integrated the Jeans equations as detailed in
Pignatelli and Galletta \shortcite{me} to reproduce the
kinematic data (see Figs.~\ref{fig:major} --
\ref{fig:minor}). The best-fit stellar M/L ratios
spheroidal components and $M_{MDO}$ are given in
Table~\ref{tab:mass}, together with their statistical
uncertainties.

For the MDO model, the error bars were obtained by
computing the reduced $\chi^2$ for the whole grid of
models, and rejecting the models with a reduced $\chi^2$
greater than the 95\% confidence level.  For the non-MDO
model, since the reduced $\chi^2$ was very high for all
of the models, we kept the models with a $\chi^2$ greater
than three times the minimum value.

\begin{table*}
\begin{center}
\begin{tabular}{ccccc}
\hline
\multicolumn{5}{c}{Model with MDO} \\

\multicolumn{1}{c}{}               &
\multicolumn{1}{c}{Bulge}               &
\multicolumn{1}{c}{Disc}               &
\multicolumn{1}{c}{Halo}               &
\multicolumn{1}{c}{MDO}            
                                        \\
Mass & 
$(1.3 \pm 0.2) \cdot 10^{10} \ \Msun$ & 
$(7.1 \pm 0.7) \cdot 10^9 \ \Msun$    &
$(2.7 \pm 0.3) \cdot 10^{10} \ \Msun$ & 
$ (7.3 \pm 2.4) \cdot 10^8 \ \Msun$ 
\\
M/L & 
$6.6 \pm 0.9$ & 
$3.3 \pm 0.3$ & 
$6.6 \pm 0.7$ & 
-- 
\\ 
\hline
\multicolumn{5}{c}{Model without MDO} \\

\multicolumn{1}{c}{}               &
\multicolumn{1}{c}{Bulge}               &
\multicolumn{1}{c}{Disc}               &
\multicolumn{1}{c}{Halo}               &
\multicolumn{1}{c}{MDO}            
                                        \\
Mass & 
$(1.5 \pm 0.2) \cdot 10^{10} \ \Msun$ & 
$(6.6 \pm 0.7) \cdot 10^9 \ \Msun$ &
$(2.8 \pm 0.3) \cdot 10^{10} \ \Msun$ &  
--  
\\
M/L & 
$7.1 \pm 1.0$ & 
$3.1 \pm 0.3$ & 
$6.8 \pm 0.6$ & 
-- 
\\ 
\hline
\end{tabular} 
\caption{Masses of the different components for the best-fit models
both with and without a central MDO. The error bars define the 95\%
confidence region in the parameter space, under the assumption that
each component has a M/L ratio constant with radius.  The mass profile
derived for the model with a central MDO is shown in
Fig.~\protect\ref{fig:mass}.}
\label{tab:mass}
\end{center}
\end{table*}

From Figure~\ref{fig:major} it is apparent that within 2.5 kpc the
galaxy can be divided into 3 different regions:
\begin{enumerate}
\renewcommand{\theenumi}{(\arabic{enumi})}
\item an inner region ($r<120\, {\rm pc}$) which features a sharp
  increase in both the stellar and gas rotation curve. This implies that
  this region is dynamically dominated by the presence of the central
  dark massive object, which also determines the central cusp in the
  velocity dispersion profile.
\item an intermediate region ($120\, {\rm pc} < r < 700\, {\rm pc}$)
  which is bulge-dominated. In this dynamically hot region the
  asymmetric drift effect dominates the stellar rotation curve, as shown
  by the large difference ($\sim 100$ km/sec, i.e. a factor of more than
  2) between the gas and the stellar velocities. The stellar rotational
  velocity reaches a plateau at about 50\% of the circular velocity.
\item an outer region ($r> 700$ pc) which is disc dominated. Here the
  stellar rotational velocity rises again, while the asymmetric drift
  effect plays a minor role. There is a probable flex point at $r \approx 900\ 
  {\rm pc}$ in the stellar velocity dispersion profile, as expected from the 
  superposition of bulge and disc populations with different kinematical
  behaviour but similar masses.
\end{enumerate}

Remarkably, these kinematic features match exactly those
expected from photometric decomposition. This is not
trivial: it implies that inside 2 kpc the light traces
the mass, thus ruling out the possibility of a sizable
presence of dark matter.

It is worthwhile to discuss in detail the differences
between the best-fit MDO and non-MDO models in
Fig.~\ref{fig:major}-\ref{fig:minor}.

First, under the assumption of M/L ratios constant with
radius for each component, the M/L values found in the
two cases are similar (see Table~\ref{tab:mass}) within
the error bars given by the fit. This reflects the fact
that in both cases the kinematics beyond $r=500$ pc
severely constrains all the parameters of the luminous
matter in both cases.

\begin{figure}
\centerline{\psfig{figure=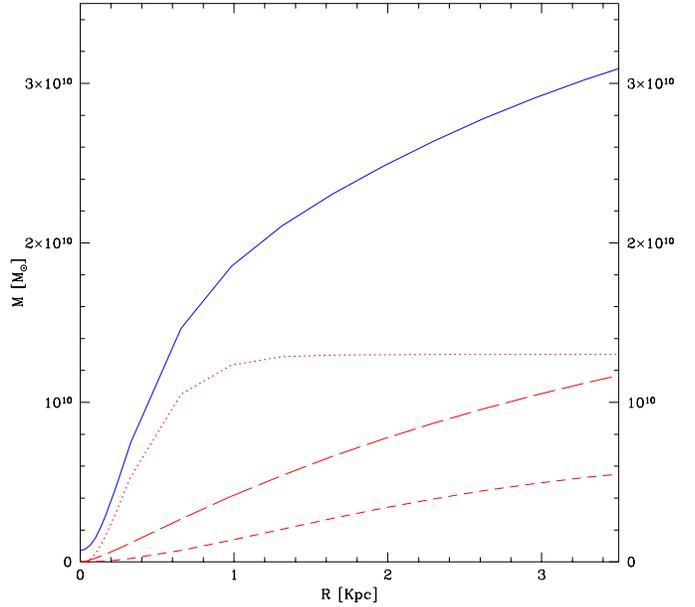,width=0.5\textwidth}}
\caption{ Luminous mass distribution of NGC 4350,
according to Table \protect\ref{tab:mass}.  We plot the
contribution of the inner exponential bulge ({\em dotted
line}), the exponential disc ({\em short-dashed line})
and the outer deVaucouleur stellar halo ({\em long-dashed
line}). The total mass is also shown as a solid line.
Note that at $r=0$ the mass is greater than zero because
of the contribution from the MDO.}
\label{fig:mass}
\end{figure}

The different panels of Fig.~\ref{fig:major} show that
the presence of the central massive object affects the
different kinematical profiles up to different radii:

\begin{itemize}

\item The velocity dispersion profile is affected inside
a ``sphere of influence'' of radius $r_{\rm BH}=
GM_{BH}/\sigma_0^2$.  This is the radius at which the
black hole is expected to increase the velocity
dispersion by a factor $\sqrt{2}$. For the observed value
of $\sigma_0$ and the inferred value of $M_{BH}$ shown in
Table~\ref{tab:mass}, we expect $r_{\rm BH}$ to be about
70 pc, which is similar to what we see in Fig.~2b.

\item While the velocity dispersion depends on the whole
mass distribution, the circular velocity is only
sensitive to the inner mass. The ``radius of influence''
in this case is the radius at which the stellar mass
equals the MDO mass, and corresponds to about 200 pc, as
can be seen in Fig.~\ref{fig:mass}. At this radius the
circular velocities of the MDO and non-MDO model differ
by a factor $\sqrt{2}$, as shown in Fig.~2a.  Beyond
$r=300$ pc part of these differences is due to the
different stellar M/L ratios (see Table~\ref{tab:mass}).
 
\item The differences between the two models in the
circular velocity are strongly amplified in the stellar
rotation curve.  For instance, at $r=300$ pc, the MDO and
non-MDO models have respective circular velocities of
about 200 and 180 km/sec.  Their asymmetric drift
corrections $\Delta V_{asymdrift}$ account respectively
for about 185 and 178 km/sec. If the rotational
velocities are computed using Eq.~(\ref{eq:asymdrift}),
these small differences are magnified as shown in
Fig.~\ref{fig:major}.

\item Finally, the MDO affects the $h_3$ moment of the
line-of-sight velocity distribution up to a similar
radius ($r\approx 300$ pc). The $h_3$ signature, together
with the gas rotation curve, are of great help in
constraining the orbital anisotropies because they rule
out strongly radial orbits in the inner part of the
galaxy.

\end{itemize}

\section{Discussion}
\label{sec:discuss}

The photometric and kinematic data along both the major
and minor axes have been explained through the superposition
of different components: an inner MDO of mass $4.9-9.7 
\cdot 10^8 \Msun$, and three stellar components
described in Table~\ref{tab:mass}, with radial mass
profiles $M(r)$ illustrated in Fig.~\ref{fig:mass}.  Each
component is discussed separately hereafter.

\subsection{The Massive Dark Object}

The rotational velocities of both gas and stars are known
to be much more efficient in constraining the mass of the
central MDO than the velocity dispersion. Models without
a central black hole fail to reproduce the shape of the
inner rotation curve (see dot-dashed line in
Fig.~\ref{fig:major}), unless unreasonable values of the
bulge mass-to-light ratio are assumed which, however,
would imply velocity dispersions much larger than those
observed.  In the inner kpc, the reduced chi-square is
$\chi \sim 1.77$ for the best-fit model with MDO, and
$\chi \sim 100$ for the model without MDO.
 
On the other hand, the errors given in Table~\ref{tab:mass}
are just statistical errors. We must consider also possible
systematic errors related to the assumptions used in the
model.

The main concern is related to different hypotheses on
the phase-space distribution function of the stellar
component.  So far, the model we presented is
axisymmetric, with an isotropic velocity
distribution. Such a model is associated with a
distribution function which only depends on the two
classical integrals of motion, i.e. energy and vertical
angular momentum: $f=f(E,L_z)$. However, strong radial
anisotropy in the centre of the galaxy results in a high
velocity dispersions, mimicking the presence of a MDO
\cite{ngc3379}.

For this specific galaxy, significant anisotropies in the
velocity distribution would certainly raise the central
velocity dispersion of the non-MDO model up to the
observed value, but: (i) the gas rotational velocity,
which is unaffected by the isotropic hypothesis, would be
too low with respect to the observations; (ii) the
stellar rotational velocity would decrease, moving away
from the observations; (iii) The radial orbits in the
central bin wound increase the $h_4$ parameter, in
contrast with that suggested by the data.  Thus, in this
galaxy the hypothesis of radial orbits can not reduce the
estimated MDO mass.

In principle, we cannot exclude the possibility that,
inside the innermost observed kinematical point, the
$M/L$ ratio of the stellar component is rising sharply
enough to mimic the presence of a massive dark
object. However, the innermost observed kinematical point
is at $R_{\rm inner} \sim 0\farcs6 = 66$ pc, and the data
requires a dynamical mass inside this point of $1.1 \cdot
10^9 \Msun$, which, compared with the observed
photometry, implies an $M/L \approx 28$ in this region,
much higher than the value $M/L \approx 11.5$ found by
Loyer et al. \shortcite{simien98}. Such a high
mass-to-light ratio seems unlikely for any given choice
of stellar population \cite{worthey}.  Adopting an
extreme $M/L \approx 12$ for the innermost region would
only reduce the best-fit $M_{MDO}$ to $ 5 \cdot 10^8
\Msun$, and the lower limit to $ 2.5 \cdot 10^8 \Msun$.

In addition, we have not forced the whole galaxy to have
the same stellar mass-to-light ratio. Instead, we allowed
different M/L ratios for different components, although
the best model fit suggests similar values for the bulge
and stellar halo. This seems a reasonable hypothesis
because it links every dynamical component to a given
star formation history. The M/L ratio of our global
stellar component increases with decreasing radius.

Finally, the hypothesis of constant flattening for each
photometric component is also a possible source of
systematic errors. Because of this assumption, we
obtained a photometric decomposition with small
statistical errors (see Table~\ref{tab:photo}), despite
the large seeing (about $2\farcs 5$) of the ground-based
observations.  If we relax this hypothesis, then it is
possible that the stellar component becomes flatter
inside the seeing radius, contributing to the rotation
shown in Fig.~\ref{fig:major} and reducing the need for a
central MDO.

We can not reproduce this kind of behaviour in a
self-consistent way with our model. We can, however, give
an estimate of the systematic errors involved, evaluating
a set of models with different ellipticities for the
bulge component, and trying to reproduce the stellar $V$
and $\sigma$ profile only in the inner $2\arcs$.  In
order to fit the inner rotation curve without MDO, all
the mass inside 2" must flattens to $\epsilon \gtrsim
0.3$. On the other hand with $\epsilon=0.3$, the central
velocity dispersion falls below the predictions for the
no MDO case with $\epsilon=0$ and does not match the
observations (cfr. Fig.~\ref{fig:major}). Under the
asssumptions of $M/L=12$ for the bulge and
$\epsilon\simeq 0.2$ within 2.5", both the inner rotation
velocity and dispersion velocity can be fitted within the
error bars, if a central dark object with $M_{MDO}\approx
1.5\cdot 10^8$ is present.

Taking into account all the above listed source of
uncertainties, the dark object mass falls in the range
$1.5\lesssim M_{MDO}\lesssim 9.7 \times 10^8$, implying
$-2 \lesssim \log (M_{MDO}/M_{\rm bulge})\lesssim -1.1$.
This wide range is substantially due to the poor
resolution of the available data.

We can now locate NGC 4350 in the $\sigma_e-M_{MDO}$
plane. Following the same prescription as in Gebhardt et
al. \shortcite{mbhsigma2}, we computed for NGC 4350 a
luminosity-weighted $\sigma_e=$180 km/s inside $R_e$.
The relationship found by Gebhardt et al.
\shortcite{mbhsigma2} would predict for this value
$6.5\times 10^7 \leq M_{MDO} \leq 9.7 \times
10^7$. Although our best estimate is a factor of 10
larger than the value expected on the basis of
$\sigma_e-M_{MDO}$ relationship, nevertheless our lower
limit is quite close to the expected range. In this
context it is worth noticing that NGC 4350 was selected
for our analysis on the basis of the high velocity
rotation of its nuclear gas.

It is interesting to compare our results with those given
in Loyer et al. \shortcite{simien98}. Adopting a single
$M/L$ ratio of 7 for the whole stellar population, they
found that the inner kinematics required a central point
mass of $M = 8\cdot 10^8\ \Msun$, a result that agrees
very well with our value of $M_{MDO}=7.3 \cdot 10^8 \
\Msun$, despite the difference in the methods
adopted. However, the evidence for a MDO was marginal.
The difference between our results and those of Loyer et
al. \shortcite{simien98} is due to the different dataset
adopted: in particular, the fact that we included both
the gas and star kinematics, and especially the $h_3$ and
$h_4$ profiles, dramatically increases the constraints we
can put on the $M_{MDO}$. This is especially true in the
inner $5\arcs$, where the shape of the line-of-sight
velocity distribution is strongly non-gaussian (see
Fig.~\ref{fig:major}), corroborating the conclusion that
the search for an MDO is much more effective if the
analysis of the higher orders of the line-of-sight
velocity distribution is included \cite{cretton99}.

\subsection{The stellar component}
\label{sec:stellar}

The photometry of this galaxy can not be described by a
simple superposition of a deVaucouleur spheroid and an
exponential disc. The addition of a third spheroidal
component is required. However, the fact that the
mass-to-light ratios that reproduce the kinematical data
for the two spheroidal components are similar suggests
that there is a single, almost spherical stellar
population; its brightness profile follows the
deVaucouleurs law only up to the inner $\approx 600$ pc,
and then starts to diverge towards a much steeper inner
surface brightness profile. This behaviour agrees with
recent HST high resolution photometric data
\cite{lauer95} showing that many ellipticals and bulges
do have inner cusps steeper than expected from the
$r^{1/4}$ law, with a luminosity density profile $I(r)
\sim r^{-\gamma}$, with $\gamma \sim 1$ for 5 pc $ <r<$ 1
kpc. In addition, Faber et al. \shortcite{faber97} have
shown that this behaviour is more evident in rapidly
rotating, disky bulges such as the one in NGC~4350.

A possible relation between the presence of a central BH
and the inner surface brightness profile has been
extensively investigated by several authors
\cite{young80,vdm98} in the framework of adiabatic black
hole growth. The effect on the photometry, however, can
only be seen in the ``sphere of influence'' of radius
$r_{\rm BH}= GM_{BH}/\sigma_0^2$, that we showed to be
about 70 pc, which is less than $1''$ at Virgo
distance. Our poor angular resolution did not allow us to
check for this effect.

It is noteworthy that the mass-to-light ratio obtained
for the spheroidal component ($(M/L)_{\rm sph} = 6.6 \pm
0.7$) is in good agreement with the relation by van der
Marel \shortcite{vdm91}, which yields $(M/L)_{sph} = 4.8
\pm 1.5$ for a galaxy of this luminosity. At the same
time, the value we found for the disc ($ (M/L)_D = 3.3
\pm 0.3$) is in agreement with Salucci \& Persic
\shortcite{sp99}, that give $(M/L)_D \approx 2.5 \pm 0.5$
at this luminosity.

We expect this difference in the $M/L$ ratio of the two
components to be visible in the colour distribution of
the galaxy. There is indeed evidence for a colour
gradient, going from $(B-V) = 1.1$ in the inner,
bulge-dominated region, to a $(B-V) \approx 0.84$ in the
intermediate region \cite{prugniel98}, which
qualitatively agrees with our results.
 
\subsection{The dark matter halo} 

In the previous sections, we purposely limited ourselves
to the inner regions of the galaxy ($ r \le 2 R_{\rm
d}$), since here the diffuse dark matter component is
clearly not required by the data.

A complete and detailed investigation of the dark matter
distribution in NGC~4350 is beyond the scope of this
paper. Nevertheless, under the hypotheses described
above, it is possible to derive a preliminary estimate on
the mass and distribution of dark matter needed to
explain the kinematical data in the outer regions.

The observed velocity \cite{fisher97} is nearly circular
and almost flat in the range $1.8\leq r \leq 3.5$ kpc. It
can be represented by a linear law $V(R) = (175 \pm 5) +
(3.6 \pm 1.2) r$ , where $r$ is expressed in kpc e $V$ in
km/sec. With the stellar distribution we derived in
Table~\ref{tab:mass}, this implies that within the last
observed kinematical point $R_{\rm last}\approx 3.5$ kpc
the dark matter amounts to $M_{DM}= 1.2 \cdot 10^{10}
\Msun$, which is 30\% of the total mass inside this
point.  The DM density should equal that of the luminous
matter at $R\approx 3$ kpc, so that, as observed in
spiral galaxies \cite{sp99}, in the inner regions the
light traces the dynamic mass.

\section{Conclusions}

From the photometric and kinematic data available in the
literature, we inferred the mass distribution for the S0
galaxy NGC~4350.  The data used to constrain the model
include: R-band photometry, rotational velocity of gas
and stars and velocity dispersion profiles along both the
major and minor axis, and the $h_3$ and $h_4$ profiles,
for a total number of 280 data points.  Our results can
be summarized as follow:

\begin{description}

\item[-] An MDO of $M \approx 1.5 - 9.7 \cdot 10^8 \Msun$
is located at the centre of the galaxy.  A lower value
for the central mass is possible if we allow for a very
sharp increase of the anisotropy in the velocity
distribution. However, this could be obtained only by
neglecting the information content in the gas kinematics
and the $h_3$ and $h_4$ parameters profiles.

\item[-] This galaxy falls in the upper envelope of the
 $M_{bulge}-M_{BH}$ or $\sigma_e-M_{BH}$ relations known so far. This
is expected, given that it has been selected for the very fast
rotation of its nuclear disc.

\item[-] The estimated stellar $M/L$ ratios ($\sim 6.6$
for the spheroid and $\sim 3.3$ for the disc) are in
agreement with the average values usually found for these
components. In order to explain the mass concentration in
the centre of NGC 4350, the stellar $M/L$ ratio has to
increase in the innermost 60 pc from the average value of
6.6 to $\sim 28$. Such a high value is, however, ruled
out by the comparison between the observed colours and
synthetic stellar population models.

\item[-] Within the hypothesis of the model, the
kinematical data between 2.5 and 3.5 kpc imply that the
DM density equalizes that of the luminous matter at
$R\approx 3$ kpc, and that the dark matter mass is 30\%
of the total within $r=3.5$ kpc.

\end{description} 

\vspace{0.2cm}\noindent
{\bf Acknowledgements.}
We thank the referee, Dr. K. Gebhardt, for helpful comments and suggestions that greatly improved the quality of the paper. 

\bibliographystyle{mn}
\bibliography{ngc4350}

\end{document}